\documentclass[sigconf]{acmart}
\usepackage{amsmath}
\usepackage{verbatim}
\usepackage{caption} 
\usepackage{subcaption}
\usepackage[utf8]{inputenc}
\usepackage[ruled,vlined]{algorithm2e}

\AtBeginDocument{%
  \providecommand\BibTeX{{%
    \normalfont B\kern-0.5em{\scshape i\kern-0.25em b}\kern-0.8em\TeX}}}

\copyrightyear{2022}
\acmYear{2022}
\setcopyright{acmcopyright}\acmConference[CHI '22]{CHI Conference on Human Factors in Computing Systems}{April 29-May 5, 2022}{New Orleans, LA, USA}
\acmBooktitle{CHI Conference on Human Factors in Computing Systems (CHI '22), April 29-May 5, 2022, New Orleans, LA, USA}
\acmPrice{15.00}
\acmDOI{10.1145/3491102.3501992}
\acmISBN{978-1-4503-9157-3/22/04}
\acmConference[CHI 2022]{ACM CHI Conference on Human Factors in Computing Systems}{April 30--May 06}{New Orleans, Louisiana, USA.}

\begin{document}

\title[Rediscovering Affordance: A Reinforcement Learning Perspective]{Rediscovering Affordance: \\A Reinforcement Learning Perspective}

\author{Yi-Chi Liao}
\email{yi-chi.liao@aalto.fi}
\orcid{0000-0002-2670-8328}
\affiliation{%
  \institution{Aalto University}
  \country{Finland}
}

\author{Kashyap Todi}
\email{kashyap.todi@gmail.com}
\orcid{0000-0002-6174-2089}
\affiliation{%
  \institution{Aalto University}
  \country{Finland}
}

\author{Aditya Acharya}
\email{a.acharya.1@bham.ac.uk}
\affiliation{
  \institution{Aalto University}
  \country{Finland}
  }
\affiliation{
  \institution{University of Birmingham}
  \country{United Kingdom}
}

\author{Antti Keurulainen}
\email{antti.keurulainen@aalto.fi}
\orcid{0000-0002-3157-1098}
\affiliation{
  \institution{Aalto University}
  \country{Finland}
}

\author{Andrew Howes}
\email{a.howes@bham.ac.uk}
\orcid{0000-0003-4251-1127}
\affiliation{
  \institution{Aalto University}
  \country{Finland}
  }
\affiliation{
  \institution{University of Birmingham}
  \country{United Kingdom}
}

\author{Antti Oulasvirta}
\email{antti.oulasvirta@aalto.fi}
\orcid{0000-0002-2498-7837}
\affiliation{%
  \institution{Aalto University}
  \country{Finland}
}

\renewcommand{\shortauthors}{Y.-C. Liao, et al.}

\begin{abstract}

Affordance refers to the perception of possible actions allowed by an object.
Despite its relevance to human--computer interaction, no existing theory explains the mechanisms that underpin affordance-formation; that is, \emph{how} affordances are discovered and adapted via interaction. 
We propose an integrative theory of affordance-formation based on the theory of reinforcement learning in cognitive sciences.
The key assumption is that users learn to associate promising motor actions to percepts via experience when reinforcement signals (success/failure) are present.
They also learn to categorize actions (e.g., ``rotating'' a dial), giving them the ability to name and reason about affordance. 
Upon encountering novel widgets, their ability to generalize these actions determines their ability to perceive affordances. 
We implement this theory in a virtual robot model, which demonstrates human-like adaptation of affordance in interactive widgets tasks.
While its predictions align with trends in human data, humans are able to adapt affordances faster, suggesting the existence of additional mechanisms. 
\end{abstract}



\maketitle

\section{Introduction}

Imagine seeing a widget for the first time.
How do you know how to interact with it?
We often ``just know'' how to do it; but this ability sometimes breaks down, and we must figure out what to do. 
Answers to this foundational problem in human--computer interaction (HCI) have built on James Gibson's concept of \emph{affordance} \cite{gibson1966senses}:
``\emph{The affordances of the environment are what it offers the animal, what it provides or furnishes, either for good or ill}'' \cite{gibson2014ecological}.
A defining aspect of affordance is its body-relativity: the tight connection between  perception and one's body.
In HCI, the concept has become part of textbooks and design guidelines \cite{norman2013design, norman1},
and has seen multiple extensions such as technological \cite{10.1145/108844.108856, 10.1145/1240624.1240799}, interactive \cite{10.1145/1274892.1274907}, and social and cultural affordances \cite{turner2005affordance, 10.1145/1028014.1028025}. 

But how do people learn affordances of the objects they interact with? 
According to the \emph{ecological perspective}, body-relative feature comparisons determine possible actions \cite{turvey1981ecological}.
For example, both the height of the steps and the length of one's legs influence the perception of whether a flight of stairs is ``climbable'' or not  \cite{warren1984perceiving}.
But body-relativity is not sufficient for HCI as it does not explain how such features are related to experience. 
For example, what if one fails to climb a particular flight of stairs; should that affordance not be changed accordingly?
We believe so. 
In contrast, the \emph{recognition} perspective, which has been studied in computer vision and AI, considers
affordance to be an act of categorization (or classification) learned via supervised learning \cite{AffordanceNet18}. 
For example, a deep neural net can be trained with images of stairs annotated as ``climbable'' or ``not climbable''. 
Upon failing to climb the stairs, images of those stairs should feed into retraining of the network, thus enabling it to see it as ``not climbable''.
Neither view explains how we discover affordances in the first place.

\begin{figure}[t!]
\centering
  \includegraphics[width=0.99\columnwidth]{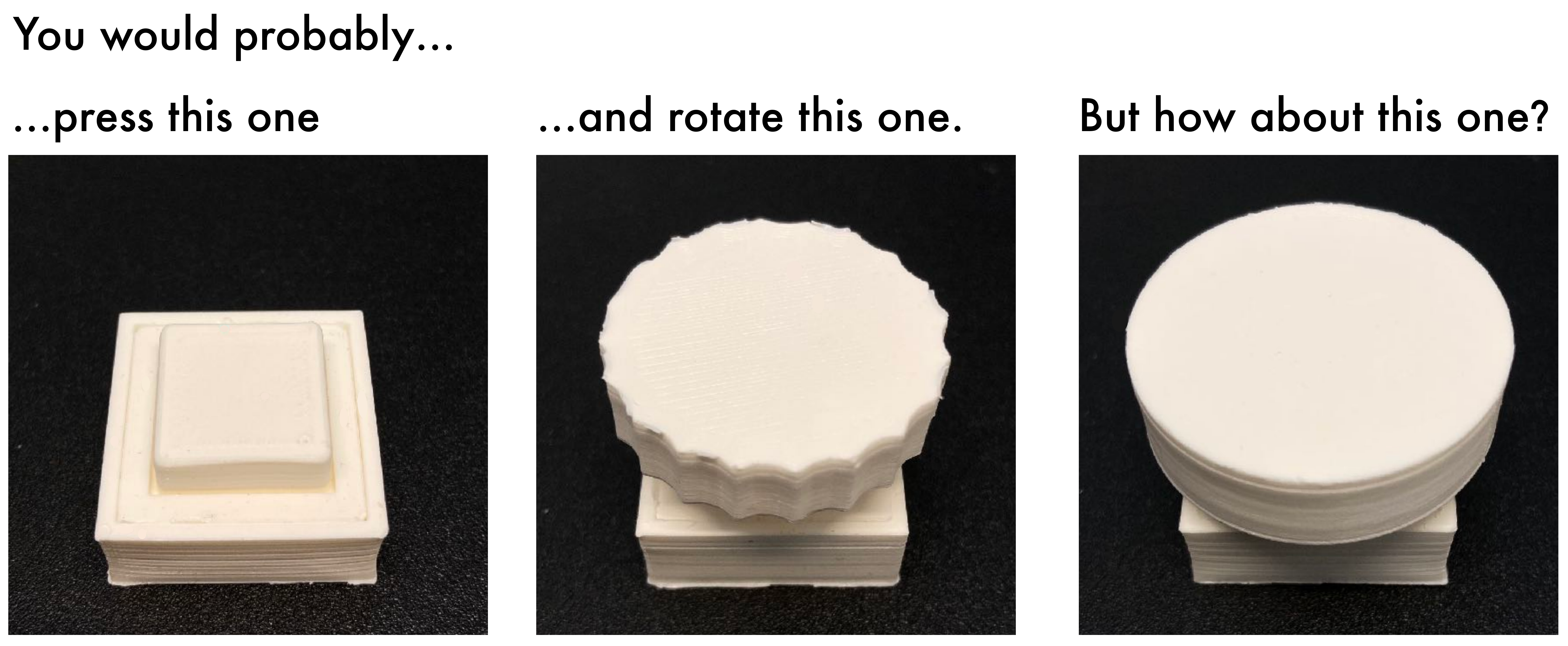}
  \caption{What action would you take with the widget on the right? This paper claims that affordances are associative percepts that map the widget's perceived features to the possible motor actions and their expected utility. The shown widgets are from our empirical study 1.} 
  \Description{A teaser image with the text ``You would probably...'' at the top; below, three physical widgets -- a square push-button, a circular dial with a chamfered edge, and a round-shaped unknown widget -- are displayed in three panels. Texts above them say ``...press this one'' (square push-button), ``...and rotate this one'' (circular dial), and ``But how about this one'' (unknown widget). Through the figure, we encourage the readers to think about what action is afforded by this unknown widget based on the prior two widgets observed.}
    ~\label{fig:teaser}
\end{figure}

This paper contributes to the understanding of affordance in HCI, 
in particular by studying an alternative cognitive mechanism that can explain affordance formation and perception.  
Furthermore, we show the application of the theory of \emph{reinforcement learning} from cognitive sciences as a tool to explain how we form and adapt affordances through experience. The theory proposed in this paper offers an integrative approach that explains how body-relative perceptions can be obtained and updated via experience. 

Our focus is on an HCI-relevant setting where interactive widgets (e.g., buttons and knobs) are presented and a user must decide how to interact with them (\autoref{fig:teaser}). 
We aim to answer two fundamental questions about affordance-formation:
\begin{enumerate}
  \item \textbf{What cognitive mechanisms underpin affordances?} We argue for two cognitive processes: 
  First, users learn associations between the expected utility of motor actions (e.g., that it is possible to turn an object with a hand) and percepts (e.g., seeing a knob).
  Second, users learn to assign categories to these motor actions (e.g., turning with hand \emph{means} ``rotating'').
  These two together allow us to perceive an action possibility \emph{with} a label. 
  Both processes have some generalization capacity: they can ``leap'' beyond previous observations, allowing us to perceive affordances with widgets we have not seen before.
  \item \textbf{How are affordances learned?} We propose that the above two types of knowledge are learned via interaction in the presence of reinforcement signals. For example, when you try to turn a widget but it does not turn, negative reinforcement helps you update your beliefs and pick a different motor action next time. 
  The theory of reinforcement learning explains how these two processes -- updating and exploring -- take place.
\end{enumerate}

In the rest of the paper, we first discuss existing theories of affordance-formation in HCI and other fields.
We then argue that reinforcement learning offers a biologically plausible and powerful explanation to the questions of affordance discovery and adaptation.
In particular, it can explain how affordance perceptions are updated in a world where we encounter novel designs all the time.
The theory achieves this without resorting to any ``special mechanism''. 
Rather, affordances are the result of everyday learning.
To understand affordance formation and perception, we present findings from two empirical studies with human participants. 
They were asked to describe or demonstrate what actions were allowed by widgets that they had not interacted with previously. 
We found evidence for different mechanisms that complement each other, enabling us to make a more accurate judgment of what actions a widget affords.
Under uncertainty (i.e., when unsure about the correct action), or when learning and discovering affordances (i.e., figuring out what is the correct ways of using a widget), participants increased the use of motion planning, simulating possible motions in their mind, which is one mechanism in reinforcement learning to predict the utility of possible actions. 
Finally, we developed a computational model of our theory that enables a virtual robot model to similarly perform affordance elicitation tasks.
We tested our model in a simulation environment, where an agent with a virtual arm and eyes interacted with different widgets.
We show that when a reinforcement signal is present, affordances could be learned interactively, as predicted.

\section{Related Work}

We review the present understanding of affordance in psychology, its relevance to design and HCI, and how it is modeled for applications in machine learning and AI. 

\subsection{Ecological Perspective in Psychology}

James Gibson spent years developing the concept of affordance.
He offered a definition in his seminal book \cite{gibson1966senses}, and concretized it later as follows \cite{gibson1977theory}:
\emph{``Subject to revision, I suggest that the affordance of anything is a specific combination of the properties of its substance and its surfaces taken with reference to an animal.''}
He later characterized it \cite{gibson2014ecological}:
\emph{``If a terrestrial surface is nearly horizontal (instead of slanted), nearly flat (instead of convex or concave), and sufficiently extended (relative to the size of the animal) and if its substance is rigid (relative to the weight of the animal), then the surface affords support.''}
To Gibson, affordance refers to opportunities to act based on features of the environment as they are presented to the animal.

Ecological psychologists elaborated on Gibson's theory.
The most agreed-on definition, endorsed by 
Heft \cite{heft1989affordances}, Michaels \cite{michaels2000information}, Reed \cite{reed1996encountering}, Stoffregen \cite{stoffregen2000affordances}, and Turvey \cite{turvey1992affordances},
is that affordances are body-relative properties of the environment that have some significance to animal’s behavior.
According to Turvey, affordances are \emph{dispositional properties} of the environment \cite{turvey1981ecological}. 
Dispositional properties are tendencies to manifest some other property in certain circumstances.
Something is ``fragile'' if it would break in certain common but possible circumstances, particularly in circumstances in which it is struck with force. 
Later, Chermo proposed \emph{relational affordance} \cite{chemero2003outline}.
In contrast to the dispositional account, relational affordances are not properties of the environment, nor of the organism, but rather the organism--environment system.

Empirical research on humans has provided evidence for the body-relative perspective. 
In particular, people are able to judge affordances reliably in various tasks \cite{warren1984perceiving,franchak2010learning,franchak2014gut}.
There is also evidence for body-relativity, such as in the stair climbing experiment \cite{warren1984perceiving} and in its extension to doorways \cite{franchak2010learning}.
People can accurately predict if a stair of dynamic height can be stepped on or not.
Although Gibson and others mostly agreed that affordance perception relies on body-relative features, it is not clear how they are acquired in the first place.
Gibson simply stated that affordances are directly perceived, and we simply \emph{pick up} the \emph{information} \cite{gibson2014ecological}.
Interestingly, evidence shows that affordances do change with practice \cite{franchak2010learning,cole2013perceiving,day2017calibration}, and some theorists have proposed that affordances are learned and developed, however without specifying how \cite{gibson2000perceptual, gibson2000ecological}.
To the best of our knowledge, our work is the first attempt to explain how this occurs through interaction. 

\subsection{Affordance in HCI and Design}

Affordance is a fundamental concept in HCI and design.
Theories related to it have been developed for years, but the process of formation and adaptation of affordances has not been explained or explored.
William Gaver first introduced affordance to HCI by using it to describe actions on technological devices \cite{gaver1991technology}.
He defined affordances as \emph{``properties of the world that are compatible with and relevant for people’s interaction. When affordances are perceptible, they offer a link between perception and action.''}
He separated affordances (the ways things can be used) from perceptible information.
Hence, he noted that ``hidden affordances'' (affordances that are not perceptible) and ``false affordances'' (wrongly perceived affordances) can exist.

Donald Norman later introduced affordance to the design field in his book\textit{The Psychology of Everyday Things} \cite{norman1988psychology}. 
In his view, affordances suggest how artifacts should be used \cite{norman1,norman2013design}:
\emph{``Affordances provide strong clues to the operations of things. [...] When affordances are taken advantage of, the user knows what to do just by looking: no picture, label or instruction is required.''}
He implied that affordances should be appropriately incorporated to guide users. 
He also elaborated that \emph{``affordance refers to the perceived and actual properties of the thing ( \cite{norman1988psychology})''}, and later rephrased affordance entirely to ``perceived affordance''.
This suggests that there are \emph{actual} affordances and \emph{perceived} affordances, and they may be different.
In contrast, according to Gibson's original definition, affordances are exclusively ``perceived action possibilities''.
\citet{mcgrenere2000affordances} later pointed out that due to the lack of a single unified understanding, HCI researchers tend to either follow Gibson's original definition \cite{ackerman1996zephyr,bers1998interactive,zhai1996influence, vicente1992ecological}, adopt Norman's view \cite{conn1995time,johnson1995comparison,nielsen1997user}, or even create their own variations \cite{mohageg1996user,shafrir1994visual,vaughan1997understanding}.
In an attempt to address the issue, \citet{mcgrenere2000affordances} proposed a framework that separated affordances from the information that specifies them.
Yet, there are problems related to the meanings of affordance and how to apply it to design \cite{problem, 10.1007/978-3-319-14956-1_17}.

More recently, researchers introduced new types of affordances to expand the concept to more complicated interactions. 
For instance, several works \cite{albrechtsen2001affordances,baerentsen2002activity,kaptelinin2014affordances} have proposed an activity-based theoretical perspective to affordances, which is concerned with the social-historical dimension of an actor’s interaction with the environment.
\citet{turner2005affordance} further classified affordances into simple affordances and complex affordances: \emph{``Simple affordance corresponds to Gibson’s original formulation, while complex affordances embody such things as history and practice.''}
Similarly, \citet{ramstead2016cultural} introduced cultural affordance, which refers to action possibility that depends on \emph{``explicit or implicit expectations, norms, conventions, and cooperative social practices.''}
While these new classifications expand the application scope of affordance, they also imply that different affordance formation processes may exist in different affordance types.
Moreover, the implications and applications of affordances in design practice remain vague.

We argue that there is a need to unveil the discovery, adaptation, and perception mechanisms of affordances.
A theory that explains affordance formation through interaction could serve as a common ground and help unify existing theories and definitions.
By offering a better understanding of the concept, it can also provide actionable means to understand and improve interfaces.

\subsection{Affordance in Machine Learning}
\label{affordance_ml}
Computational models for affordance have been presented in computer vision and robotics. 
A large number of works consider affordance detection as a combined task of recognizing both the object and the actions it allows. 
Nguyen et al. \cite{nguyen2017object} suggest a two-phase method where a deep Convolutional Neural Network (CNN) first detects objects; based on this detection, affordances are observed by a second network as a pixel-wise labeling task.
\citet{AffordanceNet18} introduce a combination of object and affordance detection by using one single deep CNN, which is trained in an end-to-end manner, instead of sequential object and affordance detection.
Chuang et al. \cite{chuang2018learning} use Graph Neural Networks (GNN) to conduct affordance reasoning from egocentric scene view and a language model based on Recurrent Neural Network (RNN) to produce explanations and consequences of actions.
The capabilities for better generalization can be improved by training low-dimensional representations of high-dimensional state representations, such as autoencoders \cite{van2016stable, finn2016deep}. 
In \citet{hamalainen2019affordance}, one type of variational autoencoder is used to produce low-dimensional affordance representation from RGB images. 
In this line of work \cite{7051290,7139369,KJELLSTROM201181, 10.1007/978-3-540-79547-6_42}, affordance detection is purely based on extracting features from images by using deep convolutional networks and supervised learning,
and there is no exploration or world models involved. 
Furthermore, the agents are not able to adapt according to the interactions with the environment.
Therefore, these models are not suitable for explaining real-world affordance perception.

A few recent works have also looked into Reinforcement Learning (RL) approaches to identify affordances via interaction with the world.
\citet{nagarajan2020learning} proposed an RL agent autonomously discovers objects and their affordances by trying out a set of pre-defined high-level actions in a 3D task environment.
Similarly, \citet{5995327} detects affordance by posing a human 3D model into certain actions on the targets and calculating the probability of success.
We are inspired by computational demonstrations like these and develop the argument that the ability to explore via trial-and-errors should be the cognitive mechanism that underpins human affordance learning.
However, these methods are limited by pre-defined actions; the agent cannot learn new actions outside the training set nor fine-tune the actions.
The assumption of having pre-trained actions is also detached from real-world experience.

Lastly, research in RL and robotic fields brought affordances to the micro-movements, that is, the small actions that an agent can take at each timestep. 
For instance, \citet{khetarpal2020can} defined affordance as the possibility of transition from a state to another desired state.
\citet{manoury2019hierarchical} trains the agent to learn the feasible primitive actions and by an intrinsically motivated exploration algorithm. 
This approach allows the agents to learn the possible actions in different states, which effectively boosts the training efficiency.
However, the affordance in these works is focused on the possibility of micro, primitive actions, which are distinct from the general notion of affordances of ``larger'' actions, such as press, grasp, sit.
Our work does not extend from this view.

Our work presents a novel framing of affordance formation based on reinforcement learning, and applies this to enable a virtual robot to learn and adapt affordances via interaction.

\section{Theory: Affordance as Reinforcement Learning}

\emph{Reinforcement learning} is a grand theory that is presently uniting cognitive neurosciences and machine learning in an effort to understand general principles of adaptive behavior \cite{gershman2017,silver2021reward,sutton2018reinforcement, todi2021adapting}.
Reinforcement learning can be defined as ``the process by which organisms learn through trial and error to predict and acquire reward'' \cite{gershman2017}.
Prior to this paper, reinforcement learning had not been developed as a psychological theory of affordance for HCI. 
For a review of other applications of this theory in HCI, see \cite{howes2018interaction}. 
In what follows, we provide a synthesis of the assumptions of the theory as they are relevant for affordance-formation, especially in HCI. 

\paragraph{Affordances are learned when reinforcement signals are provided in response to motions.}
To learn which motor command leads to the highest reinforcement signal (reward),
an organism must try out several possible motions.
This experience results in concomitant updates in predicted rewards.
For example, assume you have never encountered a rotary dial before. 
If your initial motion (e.g., push) does not lead to the expected or desired feedback, you receive a negative reinforcement signal. 
The rational response then is to avoid that motion in the future \cite{chater2009rational}. 
Even when repeating a correct motion, the exact strategy or action may be further optimized to acquire the positive reward faster and with less effort.

\paragraph{Affordance perception is guided by predicted rewards.}
The theory of reinforcement learning suggests that prediction is the key to the problem posed by delayed feedback \cite{sutton2018reinforcement}.
In order to pick an action right now, the brain must learn to \emph{predict} how good eventual outcomes that choice may lead to \cite{gershman2017}.  
The better these predictions are, the better the action that is based on them. 
In cognitive neuroscience, dopamine is identified as the transmitter of phasic signals that convey what are called reward prediction errors \cite{gershman2019believing}.
We hypothesize that the emergent role of rewards is to report the salience of perceptual cues that leads to a sequence of actions. In this sense, rewards mediate the affordance of cues that elicit motor behavior \cite{cisek2007cortical}; in much the same way that attention mediates the salience of cues in the perceptual domain \cite{itti1998model}.

\paragraph{Affordances are learned by exploring and exploiting.}
Learning affordances purely via trial and error would be highly inefficient, as there are too many possible motions to try out at random. 
According to the theory of reinforcement learning, a rational agent, after having identified a satisfactory way to perform,
will keep doing that as long as that strategy works (exploitation).
In the absence of such a strategy, or when it fails, it needs to explore other options.
But when to try out which motion?
This is the so-called \emph{exploration/exploitation dilemma} \cite{sutton2018reinforcement}.
The theory purports that an organism should pick the action that it expects to accumulate most rewards in some window of time.

\paragraph{Affordance perception generalizes to unseen instances of a category.} 
Affordance perceptions need to generalize. 
Consider, for example, seeing a beige cup with a triangle-shaped yellow handle. 
You may have not seen this particular cup before, yet you know how to grasp it. 
Somehow your experience with thousands of cups in your life transfers to this particular cup.
Reinforcement learning proposes two cognitive mechanisms for generalization: 1) generalization of policies via feature-similarity and 2) generalization via motion planning in mind. 
In the former, two objects that share similar perceptual features are associated with the same policy.
The shapes that indicate a cup, and possibly its context, are associated with grasping more than anything else. 
Such generalization can be modeled, for example, via deep neural nets \cite{lecun2015deep}, as in our computational model.
In the latter, we simulate motions with our bodies and predict their associated consequences.
In reinforcement learning theory, this is called model-based reinforcement learning \cite{gershman2017}:
a mode of reinforcement learning that is associated with better generalizability but higher effort.

\paragraph{We learn to associate categories (labels) with affordances.}
Everything that we have stated above could be applied to any animal, not just humans. 
However, affordance, as it has been treated in previous research, has almost always assumed linguistic categorization of actions \cite{ugur2009affordance, castellini2011using}.
We concur and argue that categorization is not just for sensemaking but an essential mechanism that accelerates the learning of affordances. 
It enables reasoning, social communication, and acculturation -- processes that boost the development of cognitive representations. 
The most straightforward way to categorize affordances considers the movement itself.
For example, motions, where the finger comes in contact with a surface and pushes it downwards, can be classified as ``presses''.
In machine learning terms, this is a classification problem where one needs to go from perceptual input vector to a distribution over possible labels. 
But we may also categorize actions based on similarities in \emph{consequences}. 
For example, when a finger pushes down on a keycap and a positive feedback signal (``click'' sound) is observed, the motion could also be classified as a ''press''.

\section{Study 1: Perceiving Affordances in Interactive Widgets}

To understand the contribution of different cognitive mechanisms in affordance perception and to test the hypothesis that they are adapted based on experience, we conducted two studies.
Study 1 aims to shed light on which mechanisms are present in affordance perceptions. 
More specifically, it seeks to answer two research questions:
\begin{enumerate}
    \item[] RQ1: Which internal mechanisms are employed during affordance perception? 
    \item[] RQ2: How does prior experience with particular types of widgets influence affordance perception? 
\end{enumerate}

The main idea of the study is to manipulate which widgets are experienced prior to seeing a novel widget (see \autoref{fig:study_1}).
Affordances have been previously studied in HCI and psychology via a method where participants are presented with some objects, and then asked to self-report perceived actions and feedback \cite{tiab2016understanding, warren1984perceiving, lopes2015affordance}. 
Our method follows this approach where affordance perceptions are elicited using a rating scheme.
We hypothesize that affordance perception is a complex process where multiple mechanisms can be flexibly employed.
In particular, building on previous work and our theory, we focused on three mechanisms:
\begin{itemize}
  \item \textbf{Feature Comparison}: Here, it is assumed that body-relative features drive our affordance perception, as claimed by the ecological perspective. That is, users compare features of an object (shape, size, texture, structure, etc.) to their own features (finger length, hand size, etc.) to decide what actions are allowed and what affordances are provided.
  \item \textbf{Recognition}: We use visual characteristics and details to identify affordances of objects. Here, attributes of the actor or agent are not considered that essential.
  \item \textbf{Motion Planning}: We simulate possible motor actions to find out which have the highest probability of succeeding when interacting with a widget. 
\end{itemize}

\subsection{Participants}
Twenty-six (26) participants (15 masculine, 10 feminine, 1 chose not to disclose), aged 21 to 33 ($mean = 27.41$, $s.d. = 4.85$), with varying educational backgrounds, were opportunistically recruited.
In the context of the COVID-19 pandemic, the experiment was carried out in accordance with local health and safety protocols.
Participants reported normal or corrected vision and no motor impairments.
The same set of participants also took part in the second study, presented in the next section.
Note that two participants failed to follow the instructions of Study 2, and their data was removed (see \autoref{study2_method} for more details).
In the following, we consider only the remaining twenty-four (24) participants.
Each study took under 30 minutes, and the total duration was less than an hour per person.
Participation was voluntary and under informed consent; participants were compensated with a movie voucher (approx. 12 EUR).

\subsection{Material}
\begin{figure*}[t!]
\centering
  \includegraphics[width=0.7\textwidth]{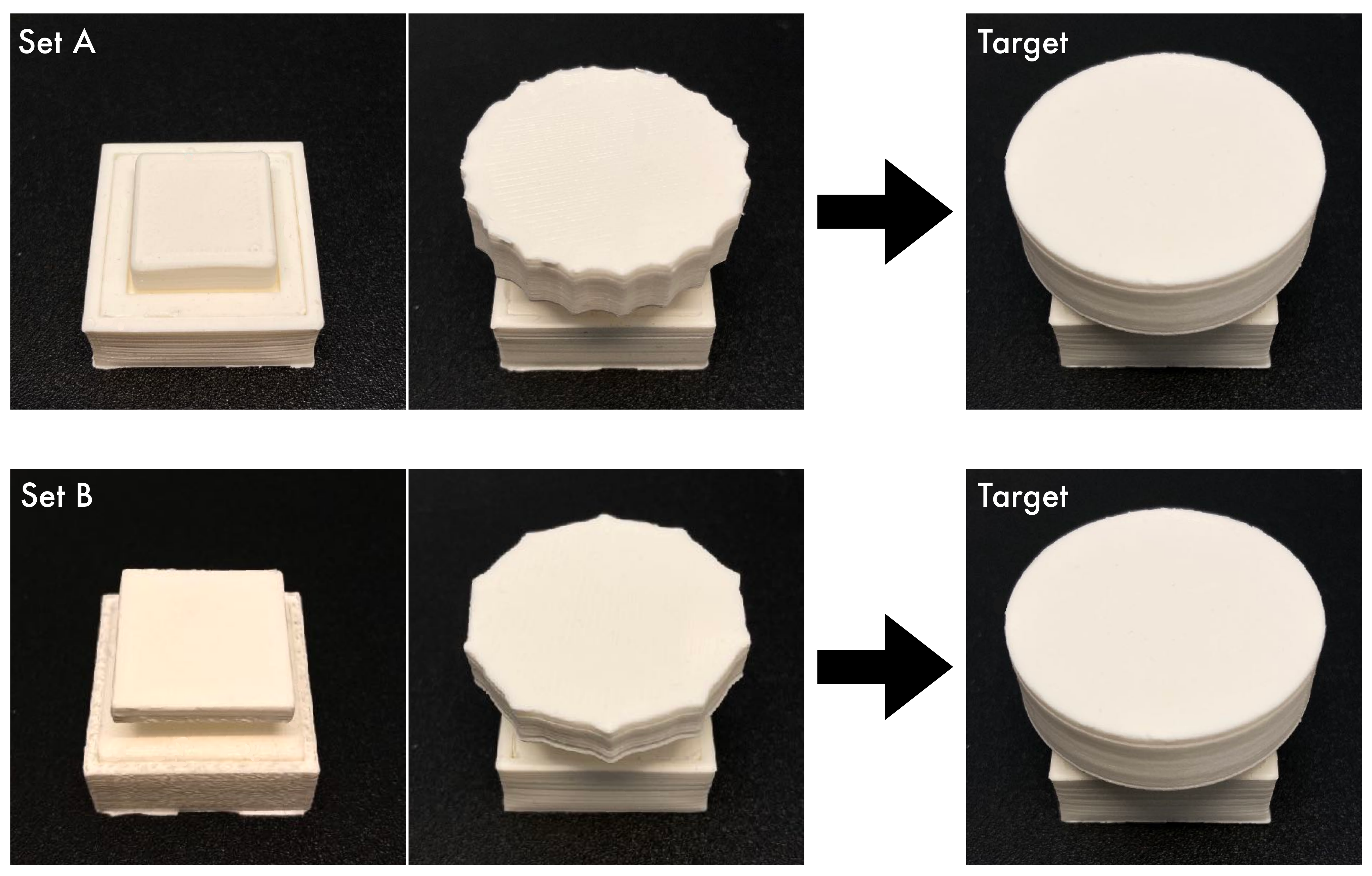}
  \caption{In Study 1, participants interacted with a set of widgets (training) and were then asked how they would interact with a novel widget (testing). We manipulated the widget sets (A or B) with which they interacted in the training phase. \textsc{Set A}: the round-shape (right-hand side) object affords rotation, and the rectangular (left-hand side) object affords pressing. \textsc{Set B}: the round-shaped (right) object affords pressing while the square object (left) affords rotation. The target object: the one which participants were asked not to interact with but just report the affordance in the ``testing phase''.} 
  \Description{The widgets and the procedure of user study 1. The details are included in the caption.}
    ~\label{fig:study_1}
\end{figure*}

Five objects were 3D-printed for the study (\autoref{fig:study_1}).
These were grouped into two sets of two interactive objects (widgets) and one additional \emph{target} object.
{\sc Set A} consisted of a rectangular pressable widget (similar to a button) and a pseudo-circular rotary widget (similar to a dial).
{\sc Set B} consisted of a rectangular rotary widget and a pseudo-circular pressable widget.
The final object (``target'') was circular and non-interactive.
Each widget consisted of a base and a handle.
The dimensions of the base was consistent across widgets ($2.3 cm \times 2.3 cm \times 0.8 cm$). 
The circular handles had a diameter of 3 cm but varied slightly in their exact shape.
The square handle had dimensions of $1.5 cm \times 1.5 cm$.
The objects were presented to the participants on a desk in front of them.

\subsection{Procedure}
The study consisted of two rounds.
Each round had a \emph{training phase} followed by a \emph{testing} phase. 

\subsubsection{Training phase:} One set of two widgets (A or B) was presented; participants were invited to interact with them to discover the right ways of operating them. 
We then asked them to provide the action(s) each object enabled to confirm that they had discovered the correct one. 
Note that we avoided mentioning the term ``affordances'' to prevent varying interpretations and bias between participants.

\subsubsection{Testing phase:} After interacting with a set of two widgets, participants were presented with the round-shaped target object (\autoref{fig:study_1}-Target), but asked to not interact with it.
Instead, they were asked to self-report the action(s) this object would allow, demonstrate these action(s) through mid-air gestures, and explain them in their own words.
A list of five possible actions\footnote{These five actions were identified during a pilot study with 5 participants. During similar tasks, these were the most frequently reported actions.} (press, rotate, pull, slide, tilt) were presented, and participants were asked to rate their affordance perception on a 7-point Likert scale for each; a rating of 1 meant that they strongly disagreed that the action was supported and 7 meant that they strongly agreed.

In the second round of the study, these two phases were repeated but with a different set of two widgets (B or A).
The presentation order of widget sets was counter-balanced between participants.

\subsubsection{Semi-structured interview:} After the two rounds, participants were asked to describe, in as much detail as possible, their process for making judgments about the perceived action(s) for the target object they encountered in both rounds. 

\subsubsection{Questionnaire:}
Next, we presented them with the three candidate mechanisms (feature comparison, recognition, motion planning), along with simple explanations for each, and asked them to rate their relevance during perception tasks on a 7-point Likert scale; a rating of 1 signified that they strongly disagreed that the mechanism played a role during their tasks, while 7 meant they strongly agreed that the mechanism played a vital role.
They were also requested to provide a rationale for their ratings.

\subsection{Result: Relevance of mechanisms for affordance perception}
We analyzed participants' ratings for each mechanism and their responses in the open-ended interviews. 
The median values for the feature comparison, recognition, and motion planning mechanisms were 4 ($IQR = 5.25$, $mean = 4.17$, $s.d. = 2.25$), 7 ($IQR = 0.75$, $mean = 6.60$, $s.d. = 0.77$), and 7 ($IQR = 1.75$, $mean = 6.21$, $s.d. = 1.03$), respectively.
A Friedman Test showed statistically significant difference between these mechanisms ($\chi^2(2) = 22.694$, $p < 0.001$). 
A post hoc analysis with Wilcoxon Signed-Rank Test was conducted with a Bonferroni correction, and the result showed that the feature comparison mechanism was reported to be the least applicable (statistically significant) compared to the recognition mechanism ($Z = -3.644$, $p < 0.001$) and the motion planning mechanism ($Z = -3.218$, $p = 0.001$).
The difference between recognition and motion planning was not statistically significant ($Z = -1.768$, $p > 0.05$).

Some participants stated that the physical features of the target object were very similar to everyday objects they had used before, so they do not need to consider the feature comparison mechanism as much:
\textit{``I have considered the size and features, but it's very quick and automatic because the shape and dimension are very common. Most of my thoughts are on considering what it is and what is the right action to do.''} (P17). 
Meanwhile, the other two mechanisms received high scores, indicating that most agreed with their relevance for affordance perception.

\textit{Feature Comparison}: Despite receiving a lower overall score, 12 participants mentioned their mental process involved making body-relative feature comparisons, indicating it was useful, but just not as much as the other mechanisms.
P14 mentioned: \textit{``The size of the top part (handle) is more or less matched with my thumb's size, so it probably is designed for pressing or tilting''}, and P20: \textit{``I think it can't be pulled because the gap between the object (the handle) and the table is too narrow for my fingers to squeeze in''}. 

\textit{Recognition}: 20 participants reported using the recognition-based approach.
Many (15) mentioned the structure and the round shape of the target object directly reminded them of the previous widget sets.
Many participants (20) mentioned that the target object reminded them of other round widgets they had encountered in daily life:
\textit{``It is very similar to a widget I have seen on a radio or a microwave.''} (P15). 
\textit{``It is a button I see a lot on different machines, so the most likely action to me is pressing-down.''} (P16). 

\textit{Motion Planning}: 
12 participants reported using motion planning: \textit{``When I see the object, I instantly imagine pushing it and rotating it (meanwhile making the gesture in mid-air), so I conclude that it has the possibilities of these motions''} (P1).
Some commented imagining motions to eliminate infeasible actions: 
\textit{``I tried to press it in imagination, but it seems not pressable. I can sort of imagine pushing and getting stuck, so I gave a lower score''.}

\subsection{Result: Changes in affordance perception during the study}

\begin{figure*}[t!]
\centering
  \includegraphics[width=0.99\textwidth]{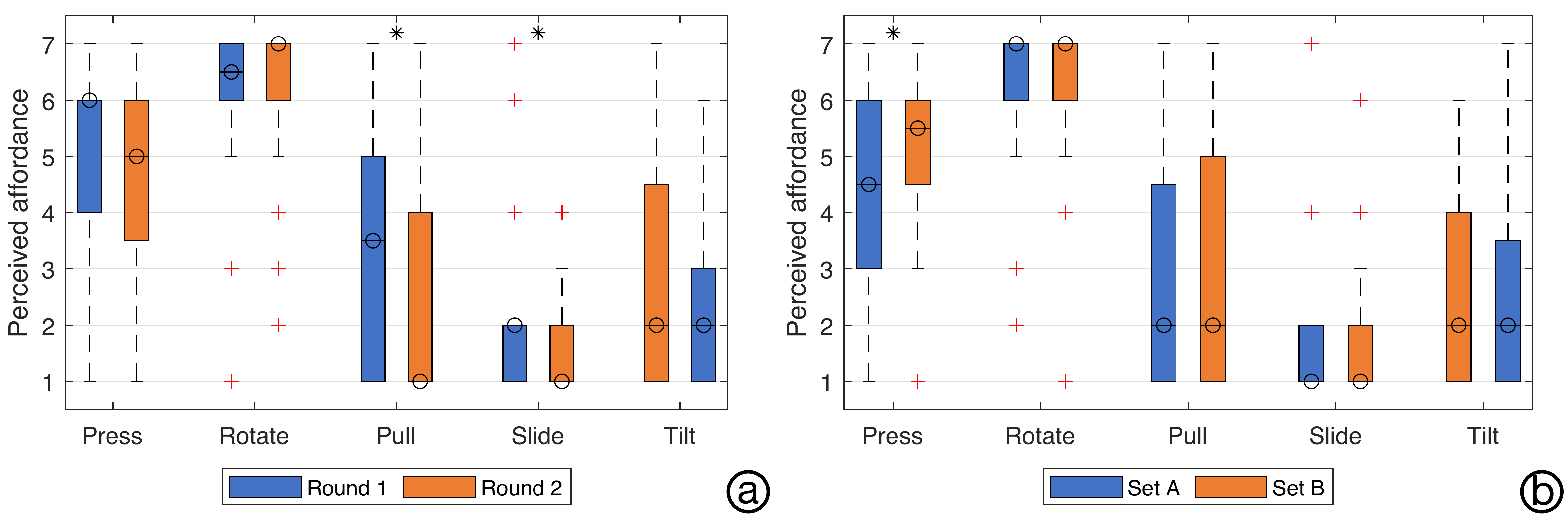}
  \caption{Study 1 shows that the reported affordances of a target widget change when the training widgets change. The red crosses mark outliers, which are defined as beyond 1.5 * IQR +/- Q3/Q1. The one-star ($*$) symbol indicates $p<0.05$ and significant difference. (a) Affordances reported for the target widget on Round 1 and Round 2. (b) Affordances depending on the preceding widget Set. } 
  \Description{The box plots show the perceived affordances in the user study 1. The detailed analysis is in section 4.5.}
    ~\label{fig:study1_result}
\end{figure*}

We looked at how perceived affordances changed as the study progressed and participants encountered different widgets.
The plots in \autoref{fig:study1_result} summarize the results.
A more detailed analysis for each action follows.

\emph{Press action}: 
For the perceived press action on the target object, Wilcoxon Signed-Rank Test showed statistically significant difference between \textsc{set A} and \textsc{set B} ($Z = -2.273$, $p < 0.05$).
The trend was consistent regardless of the presentation order of the two sets.
This shows that recent interactions with other objects (training phase) influence the affordance perception with a new object. 

During the open-ended interviews, we probed to identify possible reasons.
10 users reported changing their rating of the press perception guided by the \emph{recognition} mechanism.
That is, upon seeing an object similar to the ones they interacted with, they recalled the nearest reference image (the round-shape object in \textsc{set A} or \textsc{set B}).
P1 mentioned: \textit{``The object (target) is round, so I link it to the previous round widgets that I have tried with. If the round thing offers rotation this time, I will think this one (target) is rotatable. If the previous round thing offers press, I also tend to think maybe it (target) also can be pressed.''}
7 users attributed their change in press ratings to \emph{motion planning}. 
That is, upon seeing a similar object, their planned motions followed the previous interactions in \textsc{set A} and \textsc{set B}.
P3: \textit{``(In the second round,) I learned that the round shape can be pressable. So when I saw the last thing (target), I directly imagined and wanted to press it. That imagination (pressing it) happens only now. I did not think of it in the previous round (interacting with set A).''}

\textit{Pull and slide actions:} 
These actions had consistent ratings between {\sc Set A} and {\sc Set B}.
We further examined the ratings for these two actions based on the overall time progress (i.e., between the two rounds) regardless of the widget sets.
Wilcoxon Signed-Rank Test showed a statistically significant difference between the perceived level of ``pull'' in the first round and the second round ($Z = -3.355$, $p < 0.05$), indicating the pull perception decreased over time.
The perceived level of ``slide'' exhibited a similar trend ($Z = -2.14$, $p < 0.05$).
This shows as participants gained experience, upon not observing these actions with any of the widgets during interactions, their perception or belief about the presence of these affordances reduced over time.
In P5's words: 
\textit{``After interacting 2 objects, I still consider pulling a little, and imagine if it's possible to do. But after experiencing four widgets and learning there are not pulling there, I just stopped to believe that is an option. I don't think of this action or try to simulate it anymore.''}

\textit{Rotate and tilt actions:} There was neither a statistically significant difference between the two widget sets nor between the round for these actions.
This can likely be attributed to the target object showing a clear hint of rotation as its shape is very aligned with other rotating objects, which participants have encountered during their everyday interactions, and not showing any indication of being tiltable due to its shallow depth.

\subsection{Summary}
To answer RQ1, we observed that users employ a combination of all three mechanisms,
although motion planning and recognition tend to play more important roles.
Addressing RQ2, our statistical analysis revealed the importance of interactions for changing or adapting the perception of affordances.
From our qualitative analysis, we learned that this change could be attributed to \emph{motion planning} and \emph{recognition}.
Findings from this study offer promising evidence for the existence of our theory during affordance perception tasks. 
To summarize, study results favor the existence of our theory, and illustrate the process of how motion-planning (RL-based affordance) leads to affordance formation and adaptation through interactions. 

\section{Study 2: Adapting Affordance Perception when Widgets Change} \label{sec:study2}

\begin{figure*}[t!]
\centering
  \includegraphics[width=0.99\textwidth]{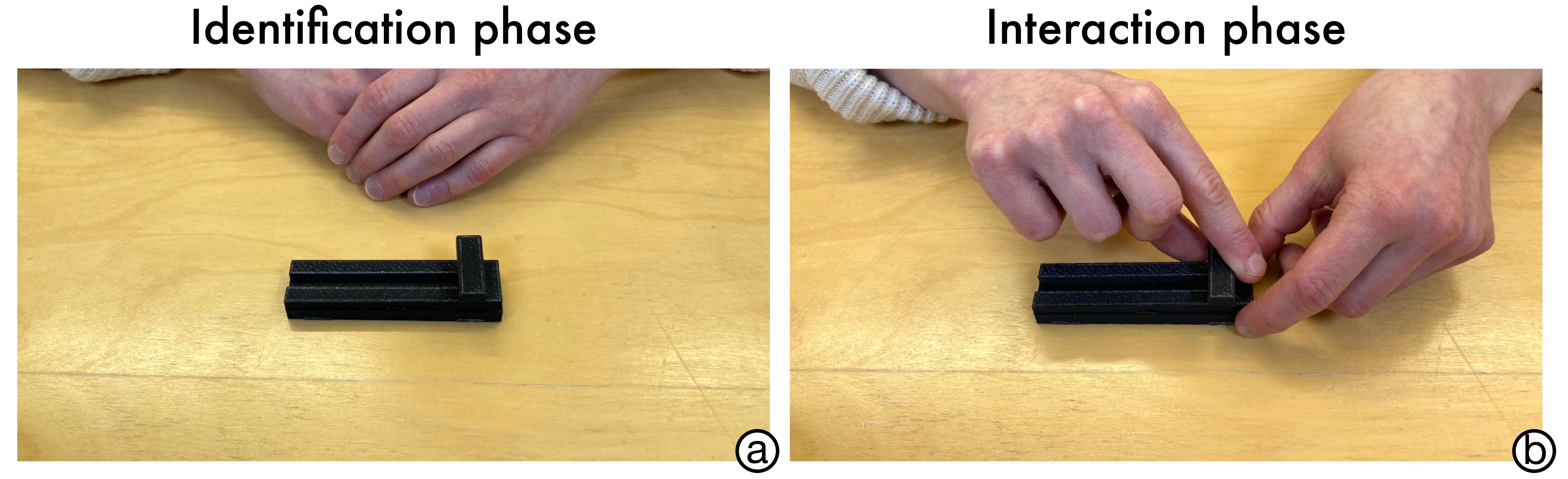}
  \caption{In Study 2, participants were shown a ``deceptive widget'' and asked to (a) identify what actions they perceived it to allow, following which they could (b) perform a single interaction with it to verify their perception.} 
    ~\label{fig:study_2}
  \Description{The widget and the procedure of user study 2. The details are included in sections 5.1 and 5.2 and the figure caption.}
\end{figure*}

The second study seeks to understand how people adapt their perception of affordances when widgets change dramatically or behave unexpectedly.
We asked participants to first identify allowed actions (affordances) on a ``deceptive widget'', which appeared to be a slider but, in fact, only allowed button-like press actions. Following this, they could perform a single interaction to verify if their perception was accurate (\autoref{fig:study_2}).
This identification and interaction could be repeated until they correctly identified the widget.
We hypothesize that users can adapt and update their affordance perception under such circumstances through motion planning and reinforcement signals (success/failure) during an interaction.
If an action leads to a failure signal (e.g., attempting to slide the handle but the handle does not move), the user updates the motion plan in the next iteration, and the detected affordance is also updated.
Conversely, if an action leads to a success signal (attempting to press the handle and successfully translating it downwards), the user determines the affordance accordingly.

\subsection{Material}
A deceptive widget that visually resembled a slider but operated like a button was 3D-printed.
The only action allowed by it was a press down on the handle.
The base of the widget was 8.5 cm $\times$ 2.5 cm $\times$ 1 cm, and the handle was 1 cm $\times$ 3 cm $\times$ 0.4 cm. 
The widget was placed in horizontal orientation, on a desk, at a distance of 15 to 20 cm from the participant.
The widget and setup is illustrated in \autoref{fig:study_2}.

\subsection{Method}
\label{study2_method}
During the study, a round consisted of two phases: \emph{identification} and \emph{interaction} (\autoref{fig:study_2}).

\subsubsection{Identification Phase:} At the start of each round, participants were asked to identify and report the most likely action allowed by the widget without making any physical contact.
If they perceived multiple actions, they could mention all of them.
In addition, they were asked to first openly describe their mental process of making the judgment.
Finally, we asked them to rate the relevance of each of the three affordance mechanisms for action identification on a 7-point Likert scale, along with follow-up questions to better understand the rationale behind their ratings.

\subsubsection{Interaction Phase:} Here, participants were asked to interact with the widget by taking the most likely action they had identified in the previous phase without making other movements.
The participants were requested to immediately inform the experimenter if they performed more than one action.
The movements were fully recorded by cameras which were set in a short distance, and all the videos were examined afterward for verification.
Two participants failed to follow the instruction and performed more than one movement in one round. 
Their data was consequently removed.

\subsubsection{Termination:} Participants could complete as many rounds as they felt necessary to confidently identify the correct action allowed by the widget.
They could also stop without identifying any action if they determined there were none allowed.

\subsection{Result: Change in Affordance Perception}

We first analyzed the perceived affordances and their change across different rounds.
On average, participants completed 3.17 rounds ($s.d. = 0.64$, $median = 3$) before stopping exploration of further possible actions (7 participants took 4 rounds, 14 took 3 rounds, and 3 took 2 rounds).
23 out of 24 participants successfully perceived the ``press'' action within 2 to 4 rounds, while one participant did not identify any possible action. 

In the first round, among the 24 participants, 22 perceived ``slide'' to be the most probable action, 1 perceived ``rotate'', and 1 ``pull''. 
Because every participant's first action led to failure, all of them decided to continue to the second round. 
Here, 7 participants considered ``press'' as the most possible action.
11 participants selected ``rotate'', 4 selected ``tilt'', and 2 chose other actions.
3 participants that correctly identified ``press'' concluded after two rounds.
21 participants attempted a third round, including 4 that had already discovered ``press'' in the second round.
9 additional participants identified ``press'' here. 
One participant failed to identify any possible actions and concluded that there was no action allowed by this object. 
All 7 participants that attempted the fourth round discovered the ``press''.

\subsection{Result: Self-reported Mechanisms}

\begin{figure}[t!]
\centering
  \includegraphics[width=0.99\columnwidth]{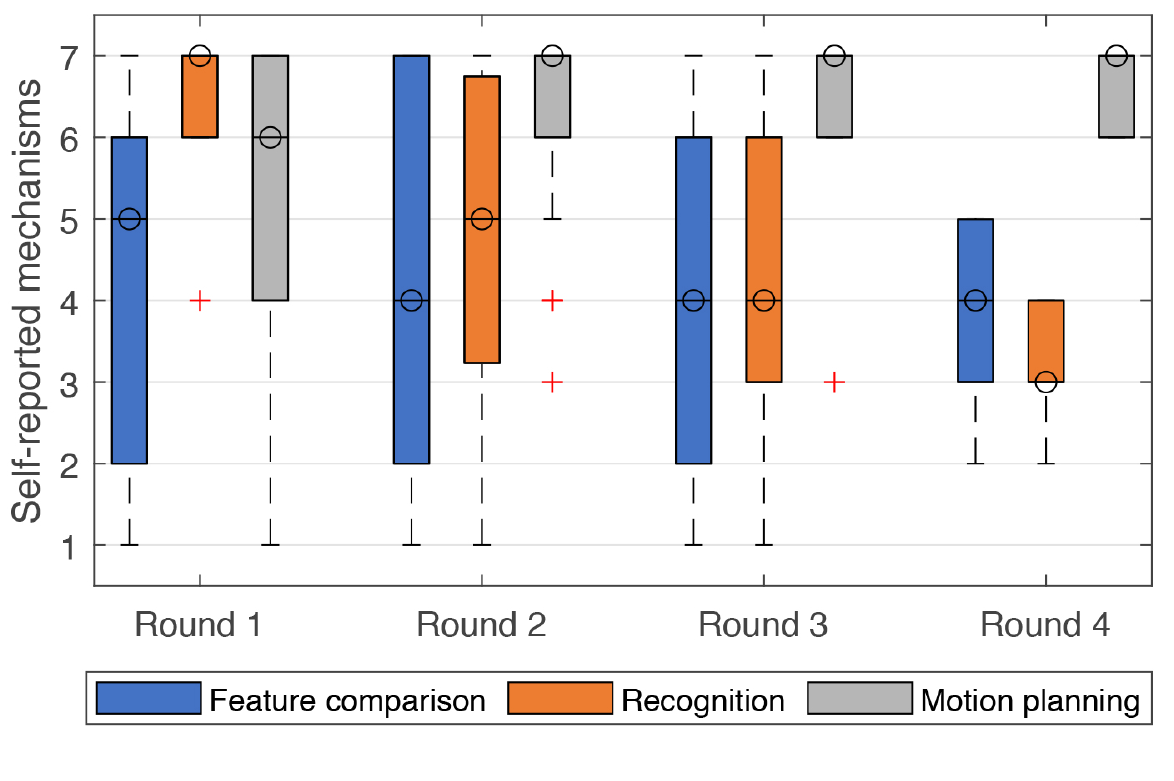}
  \caption{The relevance of each mechanism in the User Study 2. The red crosses mark outliers, which are defined as beyond 1.5 * IQR +/- Q3/Q1. Results suggest that users were initially relying on all three mechanisms but shifted to motion-planning as they gained more experience with the widget.} 
  \Description{The box plot shows the relevance of the three candidate mechanisms in different rounds.}
    ~\label{fig:study_2_result}
\end{figure}

In addition to identifying perceived actions, participants also self-reported the mechanisms they employed for identifying and adapting their affordance perception.
\autoref{fig:study_2_result} provides an overview of the results.
It is evident that the three mechanisms were similarly used in the first round.
However, once the participants interacted with the widget and noticed that their perception was inaccurate, the relevance of recognition and feature comparison consistently dissipated and motion planning gained prominence.
Eventually, motion planning was the most relevant mechanism at play when identifying and adapting perception under uncertainty.

Since only seven (7) participants reached the last round, we analyzed the data of only the first three rounds with Friedman Tests.
If there was a statistically significant difference, we used Wilcoxon Signed-Rank Tests with a Bonferroni correction for post hoc analysis.
In the first round, there was no statistically significant difference between the reported mechanisms ($\chi^2(2) = 10.204$, $p > 0.05$).
In the second round, there was a statistically significant difference between the mechanisms ($\chi^2(2) = 14.075$, $p < 0.05$).
Motion planning had the highest rating ($median = 7$, $IQR = 1$), followed by feature comparison ($median = 4$, $IQR = 5$) and recognition ($median = 5$, $IQR = 3.5$).
Post hoc analysis showed a statistically significant difference between motion planning and feature comparison ($Z = -2.623$, $p = 0.009$) and between motion planning and recognition ($Z = -2.534$, $p = 0.011$); 
the difference between feature comparison and recognition was not statistically significant.
In the third round, there was again a statistically significant difference between the mechanisms ($\chi^2(2) = 26.471$, $p < 0.001$).
Motion planning had the highest rating ($median = 7$, $IQR = 1$), followed by feature comparison ($median = 4$, $IQR = 4$) and recognition ($median = 4$, $IQR = 3$).
Post hoc analysis showed a statistically significant difference between motion planning and feature comparison ($Z = -3.208$, $p = 0.001$) and between motion planning and recognition ($Z = -3.131$, $p = 0.002$); 
the difference between  feature comparison and recognition was not statistically significant.
While we did not run statistical analysis for the data in the last round due to too few data points, we can observe a similar trend that motion planning was generally more used.

Open-ended comments shed further light on how participants applied the mechanisms during the task.
In the first round, most participants (22) reported perceiving the ``sliding'' action based on the recognition mechanism. 
As participants noted: \textit{``It is a slider''} (P5), \textit{``Reminds me of the slider on a panel to control volume''} (P20).
Similarly, many participants (15) recalled using motion-simulation as a strategy. 
As P17 said: \textit{``The motion of holding it and pushing (sliding) it along the direction just came to my mind naturally.''}
However, after failure in the first round, participants found the recognition mechanism to be less useful because the visual details of the deceptive widget did not strongly resemble objects other than a slider.
As P1 said in the second round: \textit{``After the previous fail, if I only look at the handle part, it starts looking like a widget for pulling. But the whole object still does not give me that clue what it is or what should I do.''}
P19 gave a blunt response after decreasing the recognition score from 7 to 2: \textit{``Because it doesn't work. I thought it's a slider by its look, but it isn't.''}
Users responded that the motion planning mechanism guided them to discover and adapt actions. 
P2 mentioned in round 2: \textit{``Even though it doesn't look like a button or a lever to me, I can still imagine pressing it or pulling it. It's coming not from knowing what it is, but more like I can see that motion possible.''}

\subsection{Summary}
This study assessed how users adapt their affordance perception when they are presented with novel objects that depart from their expectations and the role of different mechanisms during this adaptation process.
We observed that initial perception was guided by all three mechanisms.
However, upon failure, participants adapted quickly, enabling them to successfully identify the appropriate affordance.
This adaptation was primarily guided by motion planning and the feedback (reinforcement signals) they received during interactions.

\section{A Virtual Robot Model}
\label{sec:robot_model}

Our studies provide evidence for the theory that human perception of affordance involves reinforcement learning where the motion planning mechanism plays a key role. 
A distinct benefit of our approach over previous theories of affordance is that it can be implemented as a generative computational model that can be subjected to tasks where its affordance-related abilities are tested. 
The theory and the model are linked via the theory of reinforcement learning in machine learning \cite{sutton2018reinforcement}.
Following this methodology, we built a computational model and demonstrated it on a robotic agent, which faced a similar task as in our study 2. 
In this section, lower case \emph{reinforcement learning} refers to the cognitive mechanism while upper case \textsc{Reinforcement Learning} (or \textsc{RL}) refers to the machine learning method.
For an introduction to RL, we refer readers to \citet{sutton2018reinforcement}.

\subsection{The Interactive Widgets Task}
Before introducing our affordance model, we briefly describe the task that was used to develop the theory and the model.
Similar to our empirical studies, it is motivated by a real-world scenario where people come across unknown objects, widgets, or interfaces in their daily lives and learn or adapt how to interact with them. 

In this task, we present a virtual robot with a set of randomly selected interactive widgets.
As shown in \autoref{fig:robot1}, the agent is presented with a widget at a random location and with a randomly sampled shape and size. 
Upon successfully reaching the goal state of the widget, it receives a reward signal.
Here, the goal state refers to the correct manipulation of the widget. 
For example, a button widget has a goal state of being pressed down, a slider with a goal state of the handle being moved from left to right. 
The goal for the agent is to learn the right affordance (action allowed) through interactions with the widget.
As a novel widget might be unconventional (similar to the widget in study 2), the agent should be able to adapt its perception appropriately.

\subsection{The Virtual Robot} 
\begin{figure*}[t!]
\centering
  \includegraphics[width=0.65\textwidth]{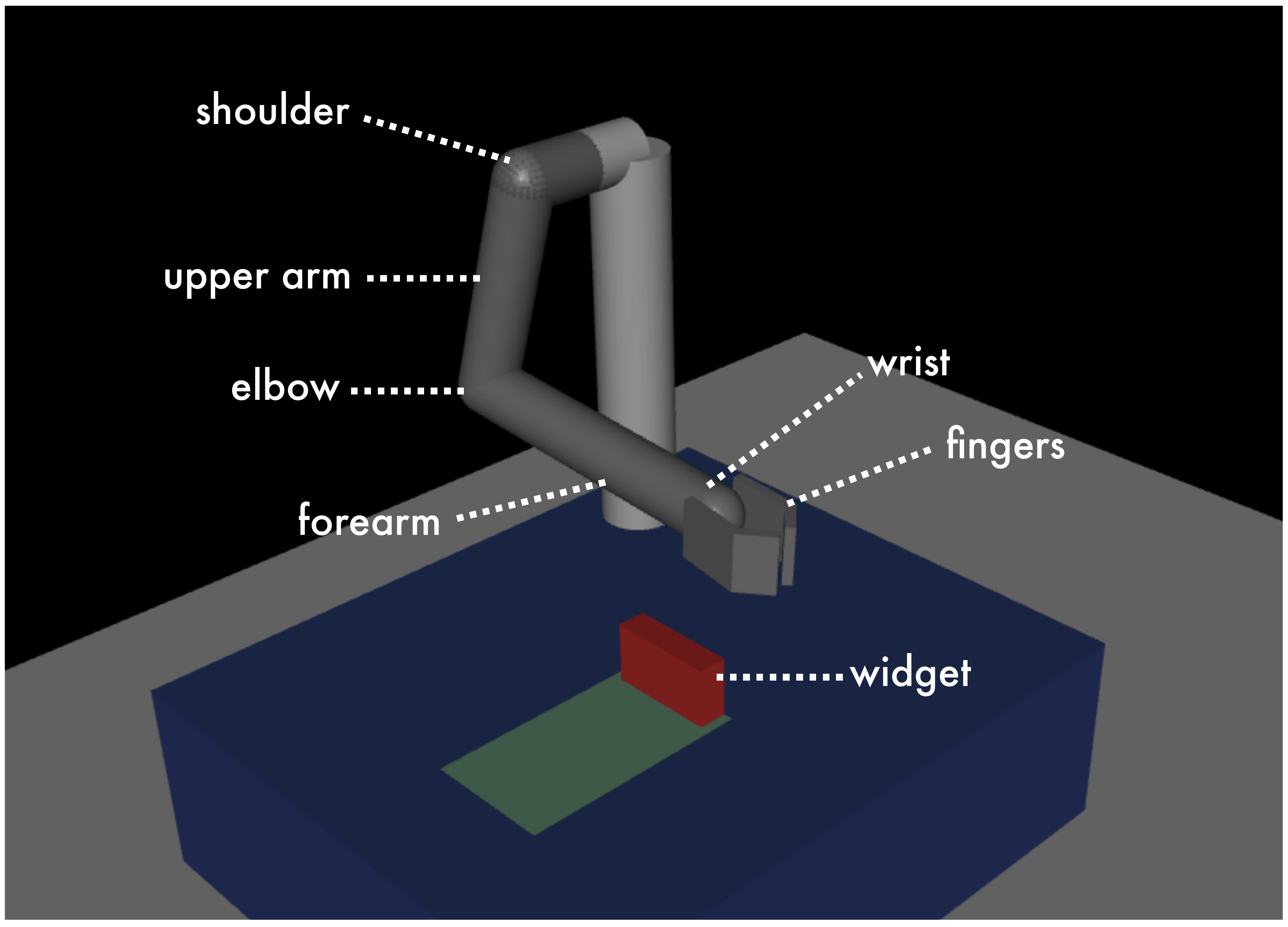}
  \caption{We develop a computational model of our affordance theory. It is implemented in the MuJoCo physical engine, and enables a virtual robot to interact with widgets.}
  \Description{The virtual robot in MuJoCo physical engine for model training and testing. The details are given in section 6.2.}
    ~\label{fig:robot1}
\end{figure*}

We implemented a virtual robot with an arm whose characteristics are similar to the human arm.
As shown in \autoref{fig:robot1}, the robot has a shoulder mounted in a static 3D position above a table, and the upper arm (24 cm long) is connected to the shoulder. 
An elbow joint connects the upper arm to the forearm (27 cm).
The other end of the forearm is the wrist and two finger-like contact points (6 cm each).
The entire robot is controlled by 7 virtual motors.
A motor can exert a force between 0 to 200 units in two possible directions.
Each motor controls one degree of freedom: two motors control the shoulder movement, one motor controls the elbow, one motor controls the rotational movement of the forearm, two motors control the wrist, and the final motor controls the grip of the two fingers. 

\subsection{Interactive Widgets}

\begin{figure*}[t!]
\centering
  \includegraphics[width=0.99\textwidth]{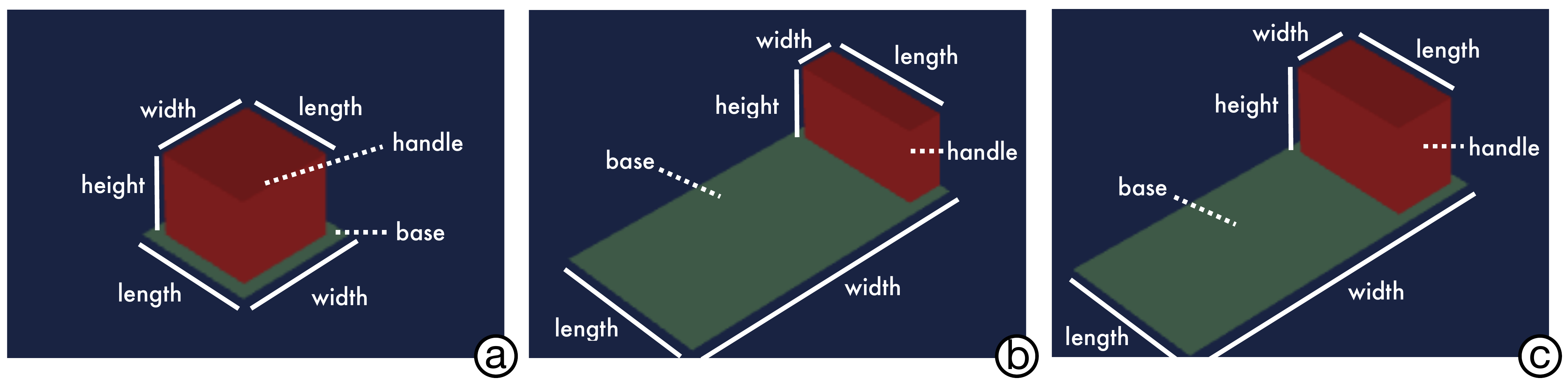}
  \caption{The virtual robot interacted with different types of widgets: (a) Button widgets; (b) Slider widgets; and (c) A deceptive widget that resembles a slider but only allows push actions similar to buttons.} 
  \Description{The widgets in MuJoCo physical engine for model training and testing. The details are given in section 6.3.}
    ~\label{fig:robot2}
\end{figure*}

We implemented two types of common widgets (\autoref{fig:robot2}) -- \emph{buttons} and \emph{sliders} -- each associated with a particular action.
A button affords the ``press'' action while a slider affords ``slide''.
In addition, we implemented a ``deceptive widget'' similar to that in our study (\autoref{sec:study2}), which appears as a slider but operates as a button.

Each widget has two parts: the \emph{handle} and the \emph{base}. 
The handle is the part the agent interacts with; we designed handles as cuboids with varying dimensions.
The base is an immovable part fixed on a table.

\begin{enumerate}
    \item Buttons (\autoref{fig:robot2}-a): 
    Every button has a handle with a square footprint (equal width and length). The exact size of the width and length is randomly selected during task trials ($\in[3 cm, 5 cm]$).
    The height of the handle is set to $3 cm$.
    The base is also square, and has a dimension 1 cm larger than the selected handle width and length to provide padding.
    
    \item Sliders (\autoref{fig:robot2}-b):
    Slider handles have a rectangular footprint, with their length being larger than their width, similar to sliders found in the real world.
    During trials, the length is $\in[4 cm, 6 cm]$ while the width is $\in[1 cm, 2 cm]$.
    The height of the handle is set to $4 cm$.
    The base of the slider is rectangular, and has a length 1 cm larger than the handle length and width 10 cm larger than the handle width.
    
    \item Deceptive Widget (\autoref{fig:robot2}-c):
    The deceptive widget appears similar to a slider; its handle is $2.5 cm (width) \times 5 cm (length) \times 4 cm (height)$, and the base is $12.5 cm (width) \times 6 cm (length)$.
    The height of the handle is set to $4 cm$.

\end{enumerate}

At the start of each trial, widget dimensions are sampled, and the origin of the widget is randomly positioned within a $5 cm \times 5 cm$ area on the table to prevent the agent from learning absolute positions.

\subsection{Modelling Assumptions}

We here describe the main ideas in how we implemented the theoretical assumptions in Section 3. 
The description requires some familiarity with RL.

First, we assume that the robot's adaptive behavior results from a solution to the Markov Decision Process (MDP) (see below). 
MDP is a mathematical framework to formulate RL problems; specifically, multi-step, sequential decision-making problems where rewards are deferred. 
A more formal definition of our MDP will be provided below.
Within this framework, we view affordances as a control problem, which maps the percepts (of the observed environment) to possible motor actions (a possible motion) through a value estimate (rewards).
This links affordance to policy models.

How to learn a policy model is a central question for \textsc{Reinforcement Learning}. 
If an action leads to a bad outcome (lower rewards), the policy model will decrease the probability of doing the same action in the future. 
If an action leads to a good outcome (higher rewards), the policy model will increase the probability of taking similar actions.
Via this process, affordances can be updated based on experience with different types of widgets. 

We further assume that people can simulate possible motions in their minds and categorize them, essentially giving them labels. 
Our implementation assumes that the agent recognizes types of actions based on similarities in movements.
For example, all motions that end with wrist-rotation movements will be classified as one type of action, and those that end with a finger poking downward will be seen as the same type of action.
In our model, we assume that these labels are given by an outside source. 
This allows implementing recognition with supervised learning.
For instance, all successful motions to activate a button are labeled as ``press''.

The proposed computational model is distinct from all the approaches that have been reviewed in \autoref{affordance_ml}. 
The agent is not detecting the affordance purely based on visual features (recognition) \cite{nguyen2017object, AffordanceNet18}. 
Instead of trying out the pre-trained actions on target objects \cite{nagarajan2020learning} or learning affordances at the level of primitive actions \cite{manoury2019hierarchical}, our agent searches for the optimal policy via exploration-and-exploitation enabled \textsc{reinforcement learning}.
Thus, the agent is able to develop novel action plans and fine-tune the learned actions for each object.

\subsection{Model Details}

We model the interaction of the agent with an object as a Markov Decision Process (MDP). 
Here, the agent takes an action $a \in A$ to interact with its environment, causing the environment's state to change from $s \in S$ to new state $s' \in S$ with a probability $T(s,a,s') = p(s' \mid s,a)$.
The reward function $R$ specifies the probability $R(s,a) = p(r \mid s,a)$ of receiving a scalar reward $r \in \mathbb{R}$ after the agent has performed an action causing a transition in the environment.
The agent acts optimally and attempts to maximize its long-term rewards.
It chooses its actions by following a policy $\pi$, which yields a probability $\pi(s,a) = p(a \mid s)$ of taking a particular action $a$ from the given state $s$.
An optimal policy $\pi^*$ maximizes the total cumulative rewards $R = E [\sum \gamma^{t-1} r^t]$ where $\gamma \in (0, 1)$ is the discount factor.
In essence, a model following an optimal policy selects the action that, in the current state of the environment, maximizes the sum of the immediate and discounted future rewards.
The MDP formulation for our model and task is as follows:

\subsubsection*{State $s$}
A state encapsulates properties of the agent and the environment, which can be observed by the agent:
\begin{enumerate}
    \item \emph{Proprioception}: The joint angles and angular velocity of all the joints of the arm.
    \item \emph{Widget dimension}: The three dimensions (width, length, and height) of the widget handle, and the two dimensions (width, length) of the base (the base is assumed to be no height).
    \item \emph{Widget position}: The 3D positions of the center of both the handle and the base.
    \item \emph{Widget velocity}: While the base is fixed, the handle is mounted on a virtual spring (for buttons) or rail (for sliders), allowing it to move.
    The velocity of the handle is encoded in the state to allow the agent to perceive the effects of its actions.
\end{enumerate}

\subsubsection*{Action $a$}
An action is a motion performed by the robot at a single time step.
As the agent has 7 joints (degrees of freedom), an action is represented as a vector with 7 values, each representing the force applied by a motor on one joint.
An interaction with a widget is composed of a sequence of such actions.

\subsubsection*{Reward $R$}
When an action $a$ is taken at a time step, the environment (in state $s$) generates a reward $r(s, a)$.
The reward is composed of three parts: distance penalty, movement penalty, and task completion reward.
The \emph{distance penalty} is set to be the distance from the center of the agent's fingers to the center of the widget handle, multiplied by a linear factor and constrained to be a value $\in [-0.01, 0]$ (the closer to the target, the less penalty received). The value is set to be relatively small to allow stable learning and avoid over-guiding.
The \emph{movement penalty} is the average joint angular velocity, multiplied by a linear factor and constrained to be $\in [-0.01, 0]$ (the faster or more you move, the more penalty received). 
This is analogous to effort or strain exerted when we make motor movements.
Finally, if the widget is successfully triggered (i.e., the button is pressed 2 cm downward or the slider is moved 4 cm toward the target direction), the agent receives a \emph{task completion reward} of value 1, otherwise 0.

\subsubsection*{Transition $T(s,a,s')$}
Taking actions results in state transitions. 
In our environment, transitions are deterministic; that is, given initial state $s$, taking an action $a$ always results in the next state $s'$ with probability 1.0.  

\subsubsection*{Discount $\lambda$} 
The discount rate is set to be 0.99 throughout the whole experiment, as it is a generally recommended value for similar robotic applications.

\subsection{Implementation}
\label{sec:model_implementation}
The task environment was implemented within the Mujoco physical simulation engine\footnote{\url{http://www.mujoco.org/}} \cite{todorov2012mujoco}, which is commonly used for robotic simulation and reinforcement learning applications. 
Our MDP agent was trained using Proximal Policy Optimization (PPO) \cite{schulman2017proximal}, a state-of-the-art \textsc{Reinforcement Learning} algorithm. For technical details and hyperparameters, refer to the supplementary material.

\subsection{Training with Common Widgets}
We trained our agent to interact with widgets that had distinct affordances.
To evaluate whether it could then adapt its affordance perception through reward signals, we tested it with a novel widget.
This widget is analogous to the widget used in our second empirical study (\autoref{sec:study2}), which investigated how humans adapt their affordance perception under uncertainty.

We trained our model to interact with the two widget types described previously -- buttons and sliders.
The training process is shown in \autoref{fig:model_eval}-a.
After training, in 1000 trials with each widget, the agent interacted with the button (by pressing it down) with a success rate of 91.3\%, and with the slider (by sliding it in the target direction) with a success rate of 94.5\%.
This offers promising evidence for the agent's ability to interact with objects through reward signals.

Next, we labeled 1000 successful trials with each widget with their corresponding affordance labels: ``press'' for the button and ``slide'' for the slider.
This dataset was used to train an affordance classifier that can assign labels to motions.
We randomly sampled 80\% of the data for the training set, 10\% for validation, and 10\% for testing. 
The motion classifier achieved 85.8\% validation accuracy and 88.1\% testing accuracy, indicating high-quality recognition.
It is challenging to achieve higher motion recognition due to some small overlap in motions that can occur during interactions with different types of widgets.
While pressing a button and sliding a slider are generally different actions, there are some motions that could successfully trigger both buttons and sliders.

\subsection{Testing with the Deceptive Widget}
After training our model to interact with common buttons and sliders, we introduced it to the deceptive widget.
Despite its resemblance to a slider, this test widget has the affordance of a button -- press actions provide positive rewards.
The agent interacted with this widget and continued learning through these interactions.
During these interactions, each action was labeled with a corresponding affordance label using the pre-trained classifier.

\subsection{Results}
\begin{figure*}[t!]
\centering
  \includegraphics[width=0.99\textwidth]{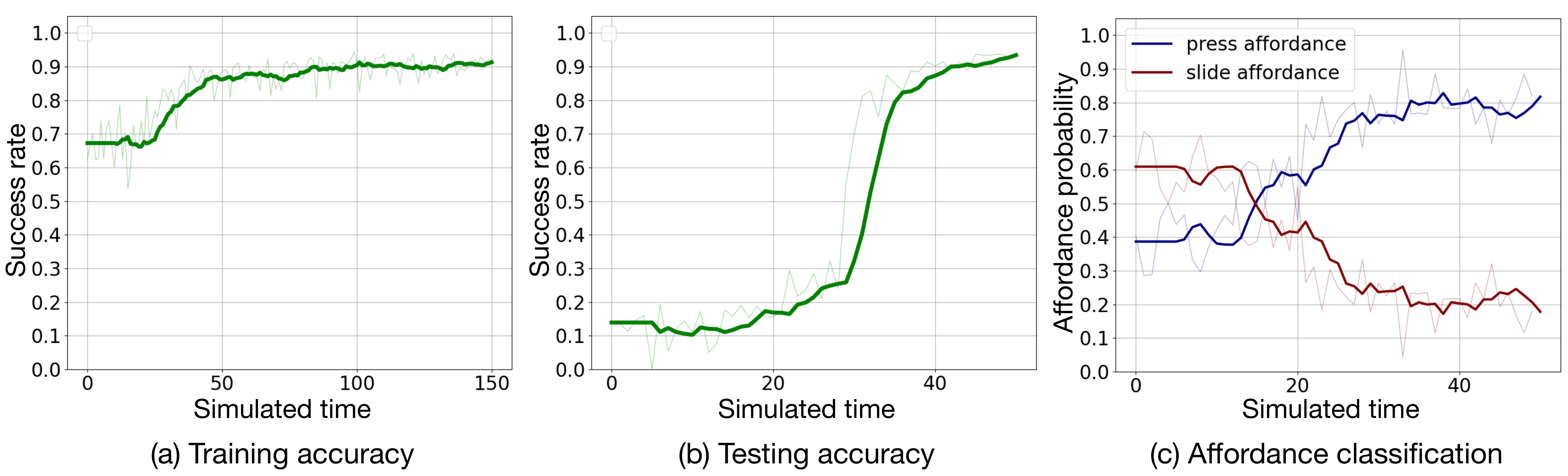}
  \caption{The virtual robot model successfully learned correct affordances by interacting with widgets. Through interactions and adaptation, it achieved a high success rate and labeled the affordance correctly. (a) The accuracy over time during the training phase with basic widgets. (b) The accuracy in the testing phase with a ``deceptive widget''. At the time 0, the testing deceptive widget was introduced. (c) Adaptation in affordance perception over time during testing.} 
  \Description{The left line chart shows the accuracy of the agent in the training phase. The middle line chart shows the accuracy in the testing phase. The right line chart shows the perceived press affordance and the perceived slide affordance.}
    ~\label{fig:model_eval}
\end{figure*}

The results are presented in the \autoref{fig:model_eval}-b and c, which shows the progress in the agent's affordance perception. 

We can see that the agent initially perceived higher sliding affordance, resulting in corresponding unsuccessful actions and low success rates.
However, through interactions, the agent was able to adapt and learned to perceive the press affordance.
As seen in the figure, after 20 simulated time units, the press affordance is dominant.
However, the agent's success rate for performing press actions does not immediately improve; this can be attributed to the difference in physical properties of this widget compared to previously learned button widgets.
Recall from our empirical study (\autoref{sec:study2}), human participants also exhibited similar behavior when perceiving affordances.
They initially indicated ``sliding'' being the primary affordance for the widget; after some interactions, they adapted and could perceive the press affordance instead.

\subsection{Summary} 
The results show that driven by \textsc{reinforcement learning}, our robot agent successfully interacts with different widgets, discovers correct affordances, provides labels for them, and adapts its perception when it encounters unfamiliar circumstances as humans do. 
While the results largely aligned with the trend in human data, the efficiency of adaptation is lower.
The potential solutions to mitigate the differences are discussed below.

\section{Discussion}

How do we learn and perceive affordances? 
Though the concept of affordance has been previously recognized and modeled using ecological and recognition approaches, the question of how affordances are formed was left unanswered. 
The paper tackles this grand question by proposing a novel theory based on reinforcement learning.
We argued that affordances are learned via trial-and-error when reinforcement signals are present, and continue to adapt in everyday interaction as we encounter different types of interfaces.
Our results are in line with the current theory in reinforcement learning, and suggest how we can generalize affordances to previously unseen interfaces. 
Through \emph{exploitation}, we can intuitively ``just see'' or perceive the right affordances and act accordingly.
When this fails, through exploration, we can resort to motion planning in mind: that is, we can simulate alternate motor actions and assess their success. 
This exploration process, which is more effortful, works similar to model-based RL in machine learning.
This result offers a neat synthesis of previous theories. In particular, the body-relative features that ecological psychology underlines are essentially our affordance percepts: they map observed features to estimated utility in operating the widget \emph{with one's own body}. Our view is also compatible with the recognition-based approach. We believe that categorization of affordances is the key to users' ability to talk and reason about affordances, which facilitates their learning. 

An interesting comparison is between our model's performance and human performance.
Human participants adapted both motion and affordance perception much faster than our robotic model.
The model took nearly 40 policy updates to achieve a 90\% success rate and perceive press affordance with a probability of 80\%.
In contrast, human participants achieved the same results within 2 to 4 trials.
Based on the analysis done, we identify three reasons that lead to this difference.
First, participants in our studies have had extensive experience of interacting with everyday objects and widgets similar to those presented in the studies; in contrast, the virtual robot had only limited experience via training.
Second, participants leveraged all three mechanisms we listed in the user study to help identify the correct affordance.
However, our model relied exclusively on motion planning, and the policy adapted to the reinforcement signal provided in each round.
Third, humans have a striking ability to learn and adapt to new environments or concepts from a limited number of examples, an ability termed \emph{meta-learning} or \emph{learning to learn} \cite{wang2021meta}.
This ability contrasts strongly with the machine learning methods used in our model, which typically require many interactions to reach a similar success rate. 
Nonetheless, our model shows a trend similar to human affordance formation and adaption. As a future extension to our model, we will consider an integrated model that can utilize multiple affordance mechanisms and will apply more advanced \textsc{reinforcement learning} techniques, such as \textsc{meta-reinforcement-learning} for more efficient learning.

Based on its ability to discover and learn affordances, our computational model has practical applications for design tasks.
Such a model can enable designers to evaluate the usability of novel design instances.
It could answer questions such as \textit{``what affordances would a user perceive when interacting with this new design?''} and \textit{``how would a user adapt their perception to a new design candidate?''}
For example, an agent could be trained to interact with a wide variety of door handles that require ``turning'' motions.
Upon encountering a novel doorknob design, the agent could reveal the level of perceived ``turn'' affordance by creating a set of motions and then classifying them, thus enabling the designer to identify whether the new design would be intuitive or not. 
As the model also adapts with experience, it could reveal how much time users might require to adapt to a novel design instance.
Furthermore, by varying the physical properties of the robot agent, such as its dimensions, degrees of freedom, or other motor capabilities, it could facilitate usability and accessibility testing for a range of user groups.

Finally, our work aims to provide a common ground for a better understanding of affordance in HCI. 
For instance, false affordances and hidden affordances \cite{gaver1991technology, wittkower2016principles} are well-known concepts that lead to poor design, but why do they exist and how to avoid them are unclear.
According to our theory, users learn to associate certain action plans to certain visual cues from past experience. 
Poor designs (with false or hidden affordances) are the ones where planned (assumed) motions do not lead to predicted reinforcement signals.
With our theory, one can further reason the quality of a design based on the planned motions and reinforcement learning.
Our theory further provides a shared formation process compatible to all the classes of affordances.
No matter whether an affordance is simple \cite{turner2005affordance} (a ball affords grasping), complex \cite{gaver1991technology,turner2005affordance} (a scrollbar on a monitor affords scrolling with a mouse), or social or cultural \cite{ramstead2016cultural}, they are all acquired via the same mechanism -- reinforcement learning. 
Lastly, our theory sparks an interesting discussion for the future: the disputed notion of natural interactions \cite{valli2008design} or natural user interfaces \cite{liu2010natural}. 
In our view, even perceiving the most \emph{natural}, \emph{intuitive} interaction possibility, such as a button pressing, is one that is \emph{discovered}, \emph{learned}, and constantly \emph{adapted} with experience.
To conclude, this paper studies the concept of affordance: a key term that was introduced to HCI decades ago. 
To date, due to the lack of a theory that explains underlying mechanisms and the formation process, the term has assumed varying interpretations and led to vague applications and implications. 
We anticipate that this paper will lead readers to rediscover affordance by shedding light on how we, as humans, rediscover and adapt our affordance perception. 
Our theory could establish a more actionable understanding of affordance in design and HCI, and our model could bring affordance from a conceptual term to a usable computational tool.

\section{Open Science}
Anonymized data from the two user studies, the virtual robot model (MuJoCo), and the RL model (Python) will be released on our project page at \url{http://userinterfaces.aalto.fi/affordance}. 
The supplementary material provides further technical details about the model implementation in \autoref{sec:model_implementation}.

\begin{acks}
This project is funded by the Department of Communications and Networking (Aalto University), Finnish Center for Artificial Intelligence (FCAI), Academy of Finland projects Human Automata (Project ID:
328813) and BAD (Project ID: 318559), and HumaneAI.
We thank John Dudley for his support with data visualization and all study participants for their time commitment and valuable insights.

\end{acks}
\balance
\bibliographystyle{ACM-Reference-Format}
\bibliography{references}

\end{document}